\documentclass[authoryear,12pt,3p]{elsarticle}
\usepackage{amsmath,bm,mathtools}
\numberwithin{equation}{section}
 \usepackage{graphicx}
\newproof{proof}{Proof}
\usepackage{lineno}

\journal{Submitted to ArXiv.org}

\begin{document}

\begin{frontmatter}



\title{Constructions for a bivariate beta distribution}


\author[io]{Ingram Olkin}
\ead{olkin@stanford.edu}
\cortext[cor1]{Corresponding author}
\address[io]{Department of Statistics, Stanford University, Stanford, CA 94305, USA}

\author[tt]{Thomas A. Trikalinos\corref{cor1}}
\ead{thomas\_trikalinos@brown.edu}
\address[tt]{Department of Health Services, Policy \& Practice, Brown University, Providence, RI 02912, USA}

\begin{abstract}
%We provide a new bivariate distribution with beta marginal distributions, positive probability over the unit square, and correlations over the full range. We discuss its extension to three or more dimensions. 
The beta distribution is a basic distribution serving several purposes. It is used to model data, and also, as a more flexible version of the uniform distribution, it serves as a prior distribution for a binomial probability. The bivariate beta distribution plays a similar role for two probabilities that have a bivariate binomial distribution. We provide a new multivariate distribution with beta marginal distributions, positive probability over the unit square, and correlations over the full range. We discuss its extension to three or more dimensions.
\end{abstract}


\end{frontmatter}

%\linenumbers

\section{Introduction}\label{sec1}

The univariate beta distribution and its bivariate extension are basic distributions that have been used to model data in various fields. For example, in population genetics, \cite{wright} showed that the beta distribution arises from a diffusion equation describing allele frequencies in finite populations, and thus the beta distribution is used to model proportions of alleles at a specific locus. By extension, a bivariate beta distribution may be appropriate to model proportions of alleles in evolutionary-related loci that are in linkage disequilibrium \citep{gianola}. Bivariate beta distributions have also been used to model drought duration and drought intensity in climate science \citep{nadarajah2007}, the proportions of diseased second premolars and molars in dentistry \citep{bibby}, tree diameter and height in forestry \citep{hafley, li, wang}, soil strength parameters (`cohesion' and `coefficient of friction') in civil engineering \citep{agrivas}, retinal image recognition measurements in biometry \citep{adell}, decisionmaker utilities in multi-attribute utility assessment \citep{libby}, and joint readership of two monthly magazines [see second example in \cite{danaher}].  

A second role for the bivariate beta distribution is that of a prior for two correlated proportions. \citet{xie} elicited a bivariate beta distribution from experts to serve as a prior distribution in the analysis of clinical trial data; and \citet{oleson} used a bivariate beta distribution as a prior distribution to correlated proportions when analyzing single-patient trials. 

The well known Dirichlet density is a multivariate generalization of the beta distribution, but it is restricted to a lower dimensional simplex. Thus it is not an appropriate model for examples such as the above. Instead, we seek a bivariate distribution with a positive probability on the unit square $(0,1)^2$, beta marginal distributions, and correlation over the full range. 

\citet*{bala} and \citet*{nelsen} discuss an array of techniques for constructing continuous bivariate distributions. We selectively outline some, to contextualize relevant literature.   

\subsection{General families}

First, one can use general families of bivariate distributions that separate the bivariate structure from the marginal distributions. Examples are the Farlie--Gumbel--Morgenstern, Plackett, Mardia, and Sarmanov families. For a fuller discussion of these and other families see \citet*{joe}, \citet*{kotz}, or \citet*{bala}. 

\subsection{Variable-in-common and transformation-based constructions}

An alternative is explicated by \citet*{libby}, who construct a multivariate generalized beta distribution starting from independent gamma variates $G_0, G_1, G_2$, with parameters $\alpha_0$ and $\beta_0$, $\alpha_1$ and $\beta_1$, and $\alpha_2$ and $\beta_2$ respectively. Then the joint density of

\begin{equation}
X=G_1/(G_1+G_0),\qquad Y=G_2/(G_2+G_0)
\label{1.00}
\end{equation}

is a generalized beta distribution with density 

\begin{equation}
f(x,y)=\frac{1}{B(\alpha_0, \alpha_1, \alpha_2)} 
 \frac{ \lambda_1^{\alpha_1}  x^{\alpha_1-1} (1-x)^{-(\alpha_1+1)} \ \ \lambda_2^{\alpha_2}  y^{\alpha_2-1} (1-y)^{-(\alpha_2+1)}} 
        {\big [ 1 + \lambda_1 x/(1-x)  + \lambda_2 y/(1-y) \big ] ^ {\alpha_0+\alpha_1+\alpha_2} }, 
\label{1.01}
\end{equation}
$\\$ for $0< x,y <1$;  $\alpha_i, \beta_i,>0$ for $i=0,1, 2$, and $\lambda_i=\beta_i/\beta_0$ for $i=1,2$. In \eqref {1.01} $B(\alpha_1,\dots,\alpha_k)=\prod\Gamma(\alpha_i)/\Gamma(\sum \alpha_i)$ is the generalized beta function.  When $\lambda_i = 1$, the density \eqref{1.01} reduces to a bivariate beta distribution with three, rather than five parameters:

\begin{equation}
f(x,y)=\frac{1}{B(\alpha_0, \alpha_1, \alpha_2)} 
 \frac{x^{\alpha_1-1} (1-x)^{\alpha_0+\alpha_2-1}   y^{\alpha_2-1} (1-y)^{\alpha_0+\alpha_1-1}}
 {(1-x y)^{\alpha_0+\alpha_1+\alpha_2}} . 
\label{1.02}
\end{equation}

\noindent \citet{jones} obtains the density \eqref{1.02} starting from a multivariate $F$ distribution. \citet*{olkin} obtain it independently using a multiplicative or logarithmically additive construction scheme analogous to that in \eqref{1.00}.   It is obvious from the construction \eqref{1.00} that $X$ and $Y$ have a positive correlation in $[0,1]$. In fact, \citet*{olkin} note that the stronger property of positive quadrant dependence holds for \eqref{1.02},  so that the probability that the bivariate random variables are simultaneously large (small) is at least as large as if they were independent.  

In addition, \citet*{nadarajah2005} note that if $U, V, W$ are beta random variates with special relations among the parameters, then {$(X=U W, \,  Y=U)$} and {$(X= U W, \, Y= V W)$} will have a bivariate beta distribution.

Finally, \citet*{arnold} construct a flexible family of bivariate beta distributions starting from five independent gamma variates $G_1$ through $G_5$, with positive shape parameters $\alpha_1$ through $\alpha_5$, respectively, and common scale parameter $1$. They define the pair $X, Y$ 

\begin{equation}
X=\frac{G_1 + G_3}{G_1+G_3+G_4+G_5},\qquad Y=\frac{G_2+G_4}{G_2+G_3+G_4+G_5},
\label{1.03}
\end{equation}
which has a density with beta distribution marginals. The joint density of $(X, Y)$ from construction \eqref{1.03} does not have a closed form and must be calculated numerically. It contains the distribution in \eqref{1.02} as a special case, namely when $\alpha_3=\alpha_4=0$. Contrary to the other constructions, it allows correlations throughout the full range. However its extension to three or more dimensions is cumbersome. 

\subsection{Generalization of existing bivariate beta densities}

\citet*{nadarajah2007} provides a slight modification to \eqref{1.02}, in which the denominator ${(1-x y)^{\alpha_0+\alpha_1+\alpha_2}}$ becomes ${(1-x y \delta)^{\alpha_0+\alpha_1+\alpha_2}}$. In the worked example the estimate of $\delta$ was close to $1$, suggesting that this parameter had a small effect in fitting the density. 

\subsection{Constructions based on order statistics}

Another construction may be via order statistics. If $X_{(1)}\leq\dots\leq X_{(n)}$ are the order statistics from a uniform distribution on $[0,1]$ then the distribution of a spacing
\begin{equation*}
w_{rs}=X_{(s)}-X_{(r)},\qquad s\geq r,
\end{equation*}
has a beta distribution,
\begin{equation*}
f(w_{rs})=\frac{w_{rs}^{d-1}(1-w_{rs})^{n-d}}{B(d,n-d)},\qquad d=s-r.
\end{equation*}
This suggests that the joint distribution of $w_{rs}$ and $w_{ts}$ is a bivariate beta distribution. We have not followed this line of inquiry but note only that it may lead to some novel results. For further discussion of order statistics see \citet*{david}.

\section{A bivariate beta distribution constructed from the Dirichlet distribution}\label{sec2}

The densities obtained with the  Farlie--Gumbel--Morgenstern, Plackett, Mardia, and Sarmanov families can have both negative and positive correlations, but, generally, over a narrow range, which can limit their usefulness as models.  Many of the aforementioned bivariate beta distributions have correlations in $[0,1]$ \citep{libby, jones, olkin, nadarajah2005, nadarajah2007}. The five parameter density introduced by \citet*{arnold} can take correlations in the full range $[-1, 1]$, which is desirable in some applications. We propose an alternative approach that has four parameters and allows correlations over the full range $[-1, 1]$.  

The marginal distributions of the Dirichlet distribution are beta distributions. Let the variates $U_{11},U_{10},U_{01}$ have the Dirichlet distribution with density
\begin{equation}
f(u_{11},u_{10},u_{01})=\frac{ u_{11}^{\alpha_{11}-1}u_{10}^{\alpha_{10}-1}u_{01}^{\alpha_{01}-1}(1-u_{11}-u_{10}-u_{01})^{\alpha_{00}-1}} {B(\bm{\alpha})},
\label{21}
\end{equation}
where $0\leq u_{ij}\leq1,\ i,j=0,1$, $u_{11}+u_{10}+u_{01}\leq1,\ \alpha_{ij}\geq0$, and $\bm{\alpha}=(\alpha_{11}, \alpha_{10}, \alpha_{01}, \alpha_{00})$. Define the additive version
\begin{equation}
X=U_{11}+U_{10},\qquad Y=U_{11}+U_{01};
\label{22}
\end{equation}
then the joint density of $X$ and $Y$ is
\begin{equation}
f(x,y)= \frac{1}{B(\bm{\alpha})} \int \limits_{\Omega} \! u_{11}^{\alpha_{11}-1} (x-u_{11})^{\alpha_{10}-1} (y-u_{11})^{\alpha_{01}-1} (1-x-y+u_{11})^{\alpha_{00}-1} \, \mathrm{d}u_{11},
\label{23} 
\end{equation} 
where
\begin{equation*}
\Omega=\{u_{11}:\max(0,x+y-1)<u_{11}<\min( x, y )\}.
\end{equation*} 

\subsection{Properties}\label{sec21}

\begin{figure}
	\includegraphics[trim=0cm 0cm 0cm 0cm, clip=true, 
					totalheight=0.95\textheight, angle=0]{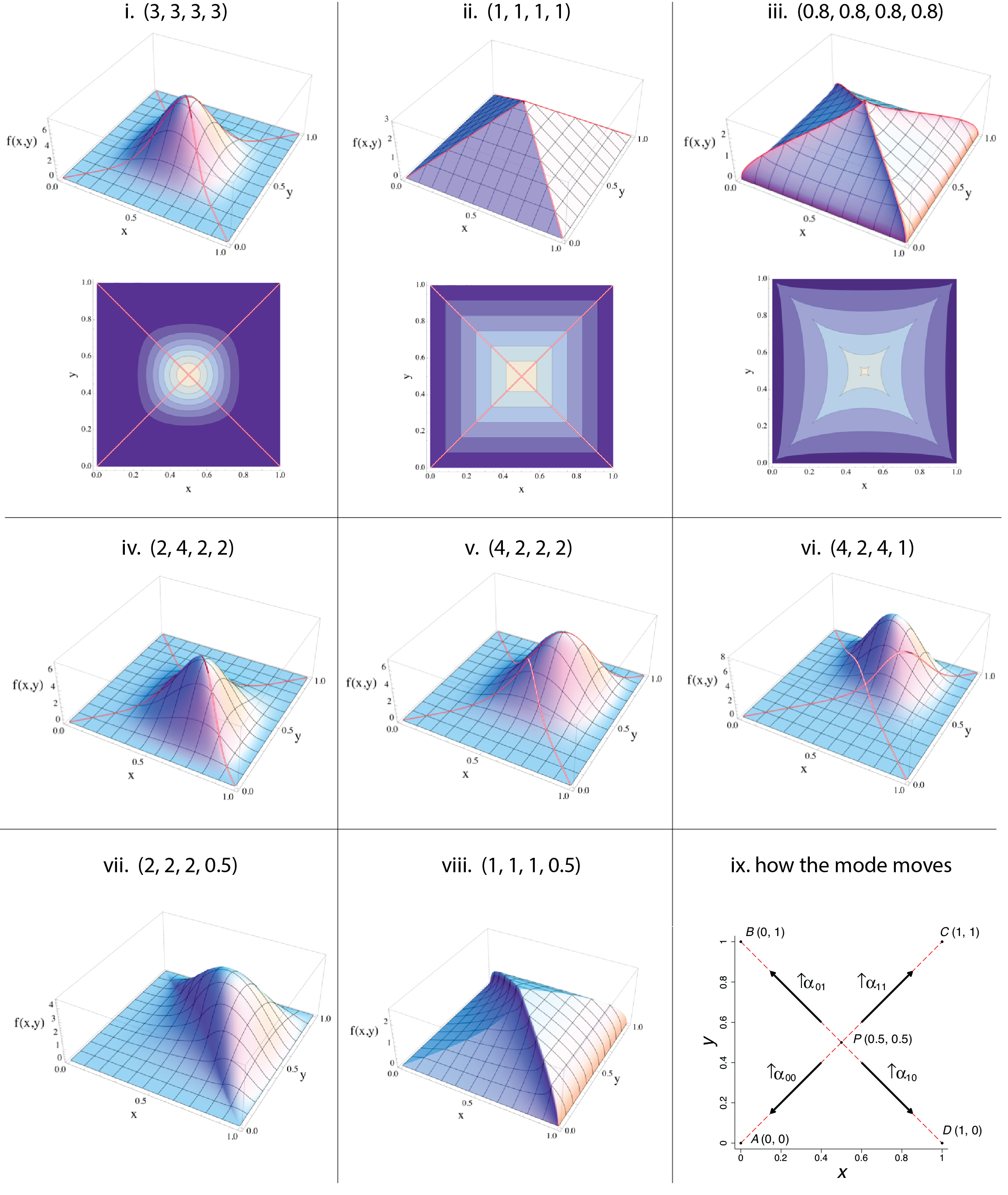}
\caption{Example densities for the bivariate beta distribution for selected parameter vector values (as listed in panels i through viii), and a schematic on the movement of the mode with increasing $\alpha_{ij}$'s, starting from all $\alpha_{ij}$ equal and $>1$, in which case the mode is at $P$ (panel ix).}
\label{fig:new_figure_v2} 		
\end{figure}

The density \eqref{23} does not have a closed form expression.  Graphs of the joint density are provided in Figure \ref{fig:new_figure_v2}. When all $\alpha_{ij}$'s are equal and larger than $1$ the mode of the distribution is at $(0.5, 0.5)$. The last panel in the figure is a schematic to build intuition on how increasing parameter values change the location of the mode.

\subsection{Notes on the density \eqref{23}}\label{sec:notes}

The density \eqref{23} is symmetric in $(x, \alpha_{10})$ and $(y, \alpha_{01})$. It can be expressed in terms of the  Appell $F_1$ generalized hypergeometric series. For example, when $x+y<1$ and $x<y$ (area $ABP$ in panel ix, Figure \ref{fig:new_figure_v2}), \eqref{23} becomes 
\begin{equation}
\label{eq24}
f(x,y)= x^{\alpha_{11}+\alpha_{10}-1} y^{\alpha_{01}-1} (1-x-y)^{\alpha_{00}-1} 
  \int_0^1 \! { \frac{ u_{11}^{\alpha_{11}-1} (1-u_{11})^{\alpha_{10}-1}  } {  (1- \frac{x}{y}u_{11})^{1-\alpha_{01}}  (1-\frac{x}{x+y-1}u_{11})^{1-\alpha_{00}}}} \, \mathrm{d}u_{11}.  
\end{equation}
The integral above is proportional to an Appell $F_1$ generalized hypergeometric function, which is defined as  
\begin{equation*}
F_1 \big (  a; b_1, b_2; c; z_1, z_2  \big )  = B(a, c-a)^{-1} \int_0^1 \! { \frac {t^{a-1} (1-t)^{c-a-1}} {(1-z_1 t)^{b_1} (1-z_2 t)^{b_2}}} \, \mathrm{d}t \text{, with } c>a>0. 
\end{equation*}
(See \citet*{bateman} for details on hypergeometric series and their integral representations). We can write \eqref{eq24} as
\begin{align*}
\label{eqABP}
f(x,y) =& B(\alpha_{11}, \alpha_{10})  x^{\alpha_{11}+\alpha_{10}-1} y^{\alpha_{01}-1} (1-x-y)^{\alpha_{00}-1} \\ 
&  \times F_1 \big (  \alpha_{11}; 1-\alpha_{01}, 1-\alpha_{00}; \alpha_{11}+\alpha_{10} ; \frac{x} {y}, \frac {x}{x+y-1}  \big ). 
\end{align*}
See the appendix for formulas expressing \eqref{23} with respect to hypergeometric functions for various lines and regions on the unit square. 

%Proceeding analogously, when $x+y<1$ and $x>y$ (area $APD$ in panel ix of the Figure) 
%
%\begin{align*}
%f(x,y) =& B(\alpha_{11}, \alpha_{01})  x^{\alpha_{10}-1}   y^{\alpha_{11}+\alpha_{01}-1} (1-x-y)^{\alpha_{00}-1} \\ 
%& \times F_1 \big (  \alpha_{11}; 1-\alpha_{10}, 1-\alpha_{00}; \alpha_{11}+\alpha_{01} ; \frac{y} {x}, \frac {y}{x+y-1}  \big ). 
%\end{align*} 
%%
%When $x+y>1$ and $x<y$ (area $BCP$ in panel ix of the Figure), the density \eqref{23} becomes  
%%
%\begin{align*}
%f(x,y) =& B(\alpha_{10}, \alpha_{00})  (1-x)^{\alpha_{01}-1} (1-y)^{\alpha_{10}+\alpha_{00}-1} (x+y-1)^{\alpha_{11}-1}  \\ 
%& \times F_1 \big (  \alpha_{00}; 1-\alpha_{11}, 1-\alpha_{01}; \alpha_{10}+\alpha_{00} ; \frac{1-y} {1-x-y}, \frac {1-y}{1-x}  \big ), 
%\end{align*} 
%%
%and when $x+y>1$ and $x>y$ (area $CDP$), it becomes 
%\begin{align*}
%f(x,y) =& B(\alpha_{01}, \alpha_{00})  (1-x)^{\alpha_{01}+\alpha_{00}-1} (1-y)^{\alpha_{10}-1} (x+y-1)^{\alpha_{11}-1}  \\ 
%& \times F_1 \big (  \alpha_{00}; 1-\alpha_{11}, 1-\alpha_{10}; \alpha_{01}+\alpha_{00} ; \frac{1-x} {1-x-y}, \frac {1-x}{1-y}  \big ). 
%\end{align*} 
%
%Finally, when $x = y$ and $x+y=1$ the density \eqref{23} has a simpler representation. It is proportional to a $_2F_1$ Gauss hypergeometric function. 

%We thank an anonymous associate editor for recognizing the connection of \eqref{23} to the hypergeometric special functions. We are indebted to Donald Richards at the Pennsylvania State University for providing an analysis. 

\subsection{Moments}\label{sec22}

The moments can be obtained explicitly as functions of the $\alpha_{ij}$ directly from the construction. The (non central) moments are
\begin{equation*}
E\,X^rY^s=E(U_{11}+U_{10})^r(U_{11}+U_{01})^s.
\end{equation*}
The respective central moments are 
\begin{equation*}
\mu_{rs}=E\,(X-\bar{X})^r(Y-\bar{Y})^s.
\end{equation*}

The case $r=0$ or $s=0$ is obtained from the marginal distribution; the case $r=s=1$ is
\begin{equation*}
E\,XY=E\,U_{11}^2+E\,U_{11}U_{10}+E\,U_{11}U_{01}+E\,U_{10}U_{01}.
\end{equation*}
Write $M=\alpha_{11}+\alpha_{10}+\alpha_{01}+\alpha_{00}$. The individual terms are
\begin{alignat*}3
E\,U_{11}^2&=\frac{B(\alpha_{11}+2,\alpha_{10},\alpha_{01},\alpha_{00})}{B(\alpha_{11},\alpha_{10},\alpha_{01},\alpha_{00})}&{}={}&\frac{\alpha_{11}(\alpha_{11}+1)}{M(M+1)},\\
E\,U_{11}U_{10}&=\frac{B(\alpha_{11}+1,\alpha_{10}+1,\alpha_{01},\alpha_{00})}{B(\alpha_{11},\alpha_{10},\alpha_{01},\alpha_{00})}&{}={}&\frac{\alpha_{11}\alpha_{10}}{M(M+1)},\\
E\,U_{11}U_{01}&=\frac{B(\alpha_{11}+1,\alpha_{10},\alpha_{01}+1,\alpha_{00})}{B(\alpha_{11},\alpha_{10},\alpha_{01},\alpha_{00})}&{}={}&\frac{\alpha_{11}\alpha_{01}}{M(M+1)},\\
E\,U_{10}U_{01}&=\frac{B(\alpha_{11},\alpha_{10}+1,\alpha_{01}+1,\alpha_{00})}{B(\alpha_{11},\alpha_{10},\alpha_{01},\alpha_{00})}&{}={}&\frac{\alpha_{10}\alpha_{01}}{M(M+1)}.
\end{alignat*}
Consequently, the covariance is
\begin{align*}
\mu_{11}=E\,XY-E\,X\,E\,Y&=\frac{\alpha_{11}(\alpha_{11}+1)+\alpha_{11}\alpha_{10}+\alpha_{11}\alpha_{01}+\alpha_{10}\alpha_{01}}{M(M+1)}-\frac{(\alpha_{11}+\alpha_{10})(\alpha_{11}+\alpha_{01})}{M^2}\\
&=\frac{\alpha_{11}\alpha_{00}-\alpha_{10}\alpha_{01}}{M(M+1)}.
\end{align*}
Also,
\begin{align*}
\mu_{20}=\text{var}(X)&=\frac{(\alpha_{11}+\alpha_{10})(\alpha_{00}+\alpha_{01})}{M^2(M+1)},\\
\mu_{02}=\text{var}(Y)&=\frac{(\alpha_{11}+\alpha_{01})(\alpha_{00}+\alpha_{10})}{M^2(M+1)},
\end{align*}
so that the correlation is
\begin{equation}
\rho=\frac{\alpha_{11}\alpha_{00}-\alpha_{10}\alpha_{01}}{\sqrt{\alpha_{1+}\alpha_{+1}\alpha_{0+}\alpha_{+0}}},
\label{24}
\end{equation}
where $\alpha_{1+}=\alpha_{11}+\alpha_{10}$, $\alpha_{+1}=\alpha_{11}+\alpha_{01}$, $\alpha_{0+}=\alpha_{00}+\alpha_{01}$, $\alpha_{+0}=\alpha_{00}+\alpha_{10}$. Note that \eqref{24} is a familiar form for $2\times2$ tables; it is obvious that $-1 < \rho < 1$, so the correlation is over the full range. Table \ref{tbl:correlations} shows correlations for selected values of parameters.

Central moments $\mu_{rs}$ of higher order can be obtained in a similar fashion.   

\begin{table}
\caption{Correlation of $X$ and $Y$ for selected parameter settings. }
\centering 
\begin{tabular}{ccccccccc}
\hline\hline  
$\alpha_{11}$ & $\alpha_{10}$ & $\alpha_{01}$ & $\alpha_{00}$  & & & & & \\
 \cline{4-9}
 & &  &10&5&2&1&.5&.1 \\
\hline
10&.1&.1&0.980&0.970&0.942&0.899&0.823&0.490\\
10&10&.1&0.490&0.394&0.266&0.182&0.112&0.000\\
10&10&.5&0.452&0.342&0.189&0.085&0.000&-0.112\\
10&10&1&0.409&0.284&0.112&0.000&-0.085&-0.182\\
10&10&2&0.333&0.189&0.000&-0.112&-0.189&-0.266\\
10&10&5&0.167&0.000&-0.189&-0.284&-0.342&-0.394\\
\hline
5&1&1&0.742&0.667&0.500&0.333&0.167&-0.076\\
5&10&1&0.284&0.167&0.000&-0.112&-0.199&-0.300\\
5&10&2&0.189&0.048&-0.141&-0.255&-0.333&-0.413\\
5&10&5&0.000&-0.167&-0.356&-0.452&-0.510&-0.563\\
\hline
2&1&1&0.576&0.500&0.333&0.167&0.000&-0.242\\
2&10&1&0.112&0.000&-0.167&-0.284&-0.378&-0.490\\
2&10&2&0.000&-0.141&-0.333&-0.452&-0.535&-0.621\\
2&10&5&-0.189&-0.356&-0.548&-0.645&-0.704&-0.757\\
\hline
1&1&1&0.409&0.333&0.167&0.000&-0.167&-0.409\\
1&10&1&0.000&-0.112&-0.284&-0.409&-0.510&-0.633\\
1&10&2&-0.112&-0.255&-0.452&-0.576&-0.663&-0.752\\
1&10&5&-0.284&-0.452&-0.645&-0.742&-0.802&-0.856\\
\hline
.5&10&.1&0.112&0.068&-0.000&-0.057&-0.119&-0.266\\
.5&10&.5&0.000&-0.085&-0.225&-0.342&-0.452&-0.621\\
.5&10&1&-0.085&-0.199&-0.378&-0.510&-0.619&-0.752\\
.5&10&2&-0.189&-0.333&-0.535&-0.663&-0.752&-0.845\\
.5&10&5&-0.342&-0.510&-0.704&-0.802&-0.861&-0.916\\
\hline
.1&10&.1&0.000&-0.040&-0.112&-0.182&-0.266&-0.490\\
.1&10&.5&-0.112&-0.201&-0.356&-0.490&-0.621&-0.823\\
.1&10&1&-0.182&-0.300&-0.490&-0.633&-0.752&-0.899\\
.1&10&2&-0.266&-0.413&-0.621&-0.752&-0.845&-0.942\\
.1&10&5&-0.394&-0.563&-0.757&-0.856&-0.916&-0.970\\[6pt]
\hline\hline
\end{tabular}
\label{tbl:correlations}
\end{table}

\subsection{Fitting a bivariate density}\label{sec23}

Given a sample $(x_n,y_n)$, $n=1,\dots,N$, we need to estimate the $\alpha_{ij}$ in order to fit a density. Denote the non-normalized sample central moments by
\begin{equation}\begin{gathered}
	m_{rs}=\sum(x_n-m_{10})^r(y_n-m_{01})^s/N, \text{ where} \\
	m_{10}=\sum x_n/N,\text{ and } m_{01}=\sum y_n/N.
\end{gathered}\label{25}
\end{equation}

We equate five central moments $\boldsymbol{\mu}=(\mu_{10}, \mu_{01}, \mu_{20}, \mu_{02}, \mu_{11})$, with the respective sample moments $\bm{m}=(m_{10}, m_{01}, m_{20}, m_{02}, m_{11})$ and solve for the $\alpha_{ij}$. The system of equations is non-linear:
\begin{equation}\begin{aligned}
m_{10}   &=(\alpha_{11}+\alpha_{10})/M,\\
m_{01}   &=(\alpha_{11}+\alpha_{01})/M,\\
m_{20}   &=(\alpha_{11}+\alpha_{10})(\alpha_{00}+\alpha_{01})/(M(M+1)), \\
m_{02}   &=(\alpha_{11}+\alpha_{01})(\alpha_{00}+\alpha_{10})/(M(M+1)), \text{and } \\
m_{11}   &=(\alpha_{11}\alpha_{00}-\alpha_{10}\alpha_{01})/(M(M+1)).
\end{aligned}\label{26}
\end{equation}

\noindent We solve  \eqref{26} by optimizing the problem  
\begin{equation}
	\begin{aligned}
	& \underset{\bm{\alpha}}{\text{minimize}} 
	& & L(\bm{\alpha}) = (\bm{m}-\boldsymbol{\mu})(\bm{m}-\boldsymbol{\mu})' \\
	& \text{subject to}  
	& & \alpha_{ij} > 0 , \text{ and} \\
	& & & \sum{\alpha_{ij}} < \text{max} \left ( \frac{m_{10}(1-m_{10})}{m_{20}} -1,\frac{m_{01}(1-m_{01})}{m_{02}} -1 \right ),
	       \; \text{for } i,j = 0,1. 
	\end{aligned}\label{27}
\end{equation}
The second constraint is an upper bound on the sum of the $\alpha_{ij}$ and follows from the marginal beta distributions. The moments match exactly when $L(\bm{\alpha}_*)=0$, or within machine precision; we call $\bm{\alpha}_*$ the solution to \eqref{27}. If  $L(\bm{\alpha}_*)>0$ the moments cannot be matched exactly.  

\noindent \textbf{Example:} \quad  Suppose $\bm{\alpha}= (4.7, 3.5, 2.1, 3.7)$.  For large $N$ 
$\\ \bm{m} \approx \boldsymbol{\mu} = (0.5857, 0.4856, 0.0162, 0.0167, 0.0034)$. Solving the problem \eqref{27} we get 
$\\ {\bm{\alpha}_*} = (4.699, 3.502, 2.101, 3.700)$, with $L(\bm{\alpha}_*)=(3.7) \; 10^{-12}$. 
To emulate a smaller sample, we perturb the sample moments to 
$\\ \bm{m}=(0.5738, 0.4647, 0.0151, 0.0170, 0.0035)$. Then 
$\\ {\bm{\alpha}_*} = (4.602, 3.639, 2.072, 4.049)$, and $L(\bm{\alpha}_*)=(1.8) \; 10^{-6}$. 
$\\$ The perturbation does not correspond to an exact solution for the system \eqref{26}, but is close enough. The results in this example are very similar if one uses additional higher order moments (up to order $3$). 

\section{Three or more dimensions}\label{sec3}

The construction \eqref{22} can be extended to $k>2$ dimensions. However it suffers from the fact that it requires $2^k-1$ components in the Dirichlet distribution. The trivariate case makes this clear. Let the random variable vector $\bm{U}=(U_{111},U_{110},U_{101},U_{011},U_{100},U_{010},U_{001})$ have a Dirichlet distribution with density 
\begin{equation}
f(\bm{u}) = C \  
		u_{111}^{\alpha_{111}-1} u_{110}^{\alpha_{110}-1} u_{101}^{\alpha_{101}-1} u_{011}^{\alpha_{011}-1}
		u_{100}^{\alpha_{100}-1} u_{010}^{\alpha_{010}-1} u_{001}^{\alpha_{001}-1} u_{000}^{\alpha_{000}-1},
\label{31}
\end{equation}
with $C=1/B(\alpha_{111},\alpha_{110},\alpha_{101},\alpha_{011},\alpha_{100},\alpha_{010},\alpha_{001},\alpha_{000})$ over the simplex $0< u_{ijk}<1$, $u_{111}+u_{110}+u_{101}+u_{011}+u_{100}+u_{010}+u_{001}=1-u_{000}>0$. Define 
\begin{equation}
\begin{matrix}
X =& U_{111} &+ U_{110} &+ U_{101} &  \  &     \    &+ U_{100} & \ & \ & , \\
Y =& U_{111} &+ U_{110} &    \     &  \  &+ U_{011} &    \            &+ U_{010} & \ & , \\
Z =& U_{111} &    \     &+ U_{101} &  \  &+ U_{011} &    \            &    \           &+ U_{001} &. \end{matrix}
\end{equation}
Then $(X, Y, Z)$ has a trivariate beta distribution in which $X, Y$, and $Z$ each have a beta distribution, and the pairs $(X, Y)$, $(X, Z)$, and $(Y, Z)$ each have a bivariate beta distribution of the form \eqref{23}.

\appendix
\section{Expression of \eqref{23} in relation to special functions}

We thank an anonymous associate editor for recognizing the connection of \eqref{23} to the hypergeometric special functions. We are indebted to Donald Richards at the Pennsylvania State University for providing an analysis, which formed the basis for the appendix. The relationship of the density \eqref{23} with the Appell $F_1$ function is as follows. \linebreak  
If $x+y<1$ and $x<y$ (area $ABP$ in the Figure, panel ix): 
\begin{align}
\label{eqABP}
f(x,y) =&  B(\alpha_{11}, \alpha_{10})  x^{\alpha_{11}+\alpha_{10}-1} y^{\alpha_{01}-1} (1-x-y)^{\alpha_{00}-1} \\ 
 \nonumber  & \times F_1 \big (  \alpha_{11}; 1-\alpha_{01}, 1-\alpha_{00}; \alpha_{11}+\alpha_{10} ; \frac{x} {y}, \frac {x}{x+y-1}  \big ). 
\intertext{If $x+y<1$ and $x>y$ (area $APD$):}
 =& B(\alpha_{11}, \alpha_{01})  x^{\alpha_{10}-1}   y^{\alpha_{11}+\alpha_{01}-1} (1-x-y)^{\alpha_{00}-1} \\ 
\nonumber & \times F_1 \big (  \alpha_{11}; 1-\alpha_{10}, 1-\alpha_{00}; \alpha_{11}+\alpha_{01} ; \frac{y} {x}, \frac {y}{x+y-1}  \big ). 
\intertext{If $x+y>1$ and $x<y$ (area $BCP$):}
 =& B(\alpha_{10}, \alpha_{00})  (1-x)^{\alpha_{01}-1} (1-y)^{\alpha_{10}+\alpha_{00}-1} (x+y-1)^{\alpha_{11}-1}  \\ 
\nonumber & \times F_1 \big (  \alpha_{00}; 1-\alpha_{11}, 1-\alpha_{01}; \alpha_{10}+\alpha_{00} ; \frac{1-y} {1-x-y}, \frac {1-y}{1-x}  \big ). 
\intertext{If $x+y>1$ and $x>y$ (area $CDP$):}
 =& B(\alpha_{01}, \alpha_{00})  (1-x)^{\alpha_{01}+\alpha_{00}-1} (1-y)^{\alpha_{10}-1} (x+y-1)^{\alpha_{11}-1}  \\ 
\nonumber & \times F_1 \big (  \alpha_{00}; 1-\alpha_{11}, 1-\alpha_{10}; \alpha_{01}+\alpha_{00} ; \frac{1-x} {1-x-y}, \frac {1-x}{1-y}  \big ). 
\end{align}

The representation is simpler across the lines $x=y$ and $x+y=1$.  Write \linebreak
$\prescript{}{2}F_1 \big( a , b; c; z  \big)= B(b, c-b)^{-1} \int_0^1 \! {t^{b-1} (1-t)^{b-c-1} (1-zt)^{-a} \, \mathrm{d}t} $ 
for the integral representation of the hypergeometric function \citep{bateman}. Then 
\begin{align}
\intertext{If $x=y <1/2$ (line $AP$ in the Figure):}
f(x,y) =& B(\alpha_{11}, \alpha_{10}+\alpha_{01}-1)  x^{\alpha_{11}+\alpha_{10}+\alpha_{01}-2} (1-2 x)^{\alpha_{00}-1} \\
\nonumber	& \times  \prescript{}{2} F_1 \big (1 - \alpha_{00},\alpha_{11}; \alpha_{11} + \alpha_{10} + \alpha_{01}-1; x / (2 x - 1) \big ). 
\intertext{If $1/2<x=y <1$ (line $PC$ in the Figure):}
=& B(\alpha_{00}, \alpha_{10}+\alpha_{01}-1)  (1-x)^{\alpha_{10}+\alpha_{01}+\alpha_{00}-2} (2 x-1)^{\alpha_{11}-1} \\
\nonumber	& \times  \prescript{}{2} F_1 \big (1 - \alpha_{11},\alpha_{00}; \alpha_{10} + \alpha_{01} + \alpha_{00}-1; (x-1) / (2 x - 1) \big ). 
\intertext{If $x=1-y <1/2$ (line $BP$ in the Figure):}
=& B(\alpha_{10}, \alpha_{11}+\alpha_{00}-1)  x^{\alpha_{11}+\alpha_{10}+\alpha_{00}-2} (1-x)^{\alpha_{01}-1} \\
\nonumber & \times  \prescript{}{2} F_1 \big (1 - \alpha_{01},\alpha_{11}+\alpha_{00}-1; \alpha_{11} + \alpha_{10} + \alpha_{00}-1; x / (1-x) \big ). 
\intertext{If $1/2<x=1-y <1$ (line $PD$ in the Figure):}
=& B(\alpha_{01}, \alpha_{11}+\alpha_{00}-1)  x^{\alpha_{10}-1} (1-x)^{\alpha_{11}+\alpha_{01}+\alpha_{00}-2} \\
\nonumber & \times  \prescript{}{2} F_1 \big (1 - \alpha_{10},\alpha_{11}+\alpha_{00}-1; \alpha_{11} + \alpha_{01} + \alpha_{00}-1; (1-x) / x \big ). 
\end{align} 

%
%
%

%\section*{References}
\bibliographystyle{model2-names}
\bibliography{beta}

%% not want to use model2-names.bst.

\end{document}